# Direct comparison of a Ca$^+$ single ion clock against a Sr optical lattice clock

Kensuke Matsubara[1], Hidekazu Hachisu[1,2], Ying Li[1], Shigeo Nagano[1], Clayton Locke[1], Asahiko Nogami[1], Masatoshi Kajita[1], Kazuhiro Hayasaka[1], Tetsuya Ido[1,2], and Mizuhiko Hosokawa[1]

[1] *National Institute of Information and Communications Technology, 4-2-1 Nukui-kitamachi, Koganei, Tokyo 184-8795 Japan*
[2] *JST-CREST, Japan Science and Technology Agency, Koganei, Tokyo, 184-8795 Japan*
[*]*matubara@nict.go.jp*

**Abstract:** Optical frequency comparison of the $^{40}$Ca$^+$ clock transition $\nu_{Ca}$ ($^2S_{1/2}$-$^2D_{5/2}$, 729nm) against the $^{87}$Sr optical lattice clock transition $\nu_{Sr}$ ($^1S_0$-$^3P_0$, 698nm) has resulted in a frequency ratio $\nu_{Ca} / \nu_{Sr}$ = 0.957 631 202 358 049 9(2 3). The rapid nature of optical comparison allowed the statistical uncertainty of frequency ratio $\nu_{Ca} / \nu_{Sr}$ to reach $1\times10^{-15}$ in only 1000s and yielded a value consistent with that calculated from separate absolute frequency measurements of $\nu_{Ca}$ using the International Atomic Time (TAI) link. The total uncertainty of the frequency ratio using optical comparison (free from microwave link uncertainties) is smaller than that obtained using absolute frequency measurement, demonstrating the advantage of optical frequency evaluation. We report the absolute frequency of $^{40}$Ca$^+$ with a systematic uncertainty 14 times smaller than our previous measurement [1].

## 1. Introduction

The most reliable means to fully evaluate the reproducibility and stability of a frequency standard is frequency comparison against independent standards. Comparison also enables the evaluation of systematic shifts by investigating the dependence of the frequency on various experimental parameters. Optical frequency standards have a significant advantage over microwave frequency standards (eg. Cs fountain clocks) in the speed of the comparison, requiring less than 1000 seconds instead of more than six hours to evaluate a fractional frequency difference with ~$10^{-16}$ uncertainty [2]. Rapid optical evaluation has enabled comparison of the frequencies of two single-ion optical clocks in same laboratory at the $10^{-18}$ level [3]. Using fiber transfer techniques [4,5] optical clocks located in distant laboratories can be compared, for example, the reproducibility of optical lattice clocks has been measured at the $10^{-16}$ level of uncertainty [6, 7]. Furthermore, an optical frequency comb can be employed as a bridge enabling the measurement of the relative instabilities of standards based on different atomic transitions. Such frequency ratio comparisons can yield information on possible temporal variations of fundamental constants [8].

In this paper, we report a frequency comparison of a $^{40}Ca^+$ $^2S_{1/2}$-$^2D_{5/2}$ single-ion clock transition against the $^{87}Sr$ $^1S_0$-$^3P_0$ lattice clock transition. The $^{40}Ca^+$ clock in this work has exceptional fractional frequency instability of parts in $10^{-16}$. Neutral strontium and calcium ions are popular as quantum absorbers in the realization of optical clocks; worldwide five neutral strontium lattice clocks are already in operation [9-13], and calcium ions are ideal for portable optical clocks due to the availability of laser-diodes at all required wavelengths. $^{40}Ca^+$ clocks are being developed at NICT [1], Univ. of Innsbruck [14], and WIPM [15].

A further reason calcium ions are gaining popularity is that they can also be used as a logic ion in quantum logic gate spectroscopy, where the logic ion acts as a coolant and state indicator of the spectroscopy ion [16] in a linear trap apparatus. A numerical simulation [17] indicates that using calcium ions as a coolant in an $Al^+$ clock has an ability similar or even better than $Be^+$ [18] or $Mg^+$ [3] that were used to demonstrate unprecedented accuracy. The relatively high sensitivity of the $Ca^+$ ion clock transition can be used as a meter of the electromagnetic field at the exact trap center and can be used to estimate possible systematic shifts of the spectroscopy ion due to electromagnetic fluctuations.

With such interest and development in these two optical standards around world it is imperative that we establish a benchmark frequency ratio with high precision.

**2. Single calcium ion clock**

The experimental setup for the $Ca^+$ clock has been improved since it was reported in [1]; firstly, an increase in production efficiency of $^{40}Ca^+$ ions has been achieved by a photo-ionization process using lasers of wavelength 423 nm ($^1S_0 \rightarrow {}^1P_1$ transition) and 374 nm ($^1P_1 \rightarrow$ ionization). Secondly, a magnetic shield has been installed on the vacuum chamber that has reduced by more than 20 times stray ac fields that previously limited the spectral width to 300 Hz. Thirdly, mechanical shutters and acousto-optic modulator (AOM) shutters have been installed to avoid coupling of cooling-laser light during the 40 ms interrogation period. Further optimization of the clock laser [19] has reduced its spectral width to less than 5 Hz, and a noise cancellation technique [20] implemented on the 40 m of optical fiber between it and the ion trap. These improvements have resulted in an observed spectral width of the clock transition of 30 Hz, as shown in Fig. 1.

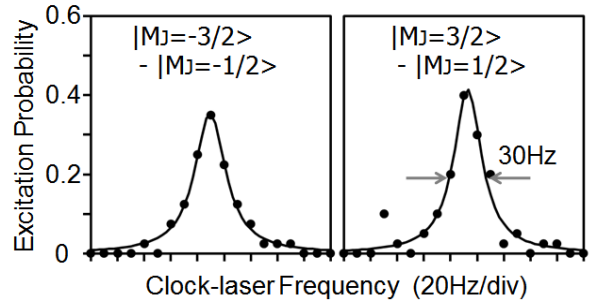

Fig.1 Spectra of $^{40}Ca^+$ clock transition. Each point of the excitation probability resulted from 40 measurements of 40 ms duration interrogation of the clock laser. The FWHM of the clock transition is 30 Hz.

The systematic shifts and the respective uncertainties to determine the $Ca^+$ clock-transition frequency are shown in Table 1. For the evaluation of the shift due to black body radiation we built an identical ion-trap equipped with platinum resistance thermometers to measure the temperature of electrodes at various positions and found when in operation the ring electrode to be at (305.4 ±2) K and the end cap electrodes (299 ± 2) K. Estimating the temperature as seen by the ion as an appropriate ratio of the two we then calculate the black body radiation shift to be 0.39 (05) Hz [21]. We estimate a gravitational shift of 3.4 (0.1) Hz from GPS-based elevation measurements, with an uncertainty of the effective geoid height of

±1 m comprising of a measurement uncertainty of ~10 cm in the geoid surface, in addition to tidal movement and the recent earthquakes. From knowledge of the magnetic field bias of 3 µT we estimate the quadratic Zeeman shift in the clock transition to be 0 (0.1) Hz. We estimate the ion temperature of 2 mK from the intensity of the motional sidebands, corresponding to a quadratic Doppler shift of 2 (10) mHz. We fully reduced the correlation of fluorescence intensity to the RF phase of the trap. To investigate possible time-dilation shifts excess voltage was deliberately applied to the micromotion compensation electrode causing large correlation between fluorescence intensity and the RF phase of the trap. No time-dilation frequency shifts larger than the measurement uncertainty were detected when using the Sr lattice clock as a reference.

In order to evaluate the electric quadrupole shift the dependence of the resonant frequencies on the upper magnetic sublevels was investigated [22]. To do this we operate an interleaved measurement between transitions having $\Delta M_J = 0$ ($|^2S_{1/2} \ M_J = \pm 1/2 >$ - $|^2D_{5/2} \ M_J = \pm 1/2>$) and transitions having $\Delta M_J = 2$ ($|^2S_{1/2} \ M_J = \pm 1/2 >$ - $|^2D_{5/2} \ M_J = \pm 5/2>$). We measure a frequency difference due to the quadrupole shift of 0.9 Hz, from which we calculate a quadrupole shift of 0.1 Hz for the normally used transition having $\Delta M_J = 1$ ($|^2S_{1/2} \ M_J = \pm 1/2 >$ - $|^2D_{5/2} \ M_J = \pm 3/2>$ ). We experimentally confirmed that quadrupole shifts of six transitions varied according to the theoretical ratio when we deliberately imposed a DC electric field. When the maximum DC field was applied (limited by requiring the ion remains stably trapped) the maximal difference between the $\Delta M_J = 0$ and the $\Delta M_J = 2$ transitions were 4.0Hz, which corresponds to the shift of 0.4 Hz for the $\Delta M_J = 1$ transition. Thus, the magnitude and uncertainty of the quadrupole shift for the $\Delta M_J = 1$ transition are estimated to be 0.1(0.3) Hz so that maximum shift of the stably trapped ion is 0.4 Hz. In addition, there was no frequency difference within measurement uncertainty between the average frequency of the three transitions [23] and the $|^2S_{1/2} \ M_J = \pm 1/2 > - |^2D_{5/2} \ M_J = \pm 3/2>$ transition.

Potential ac Stark shift due to 397-nm light (used for cooling and detection) unintentionally coupling into the optical fiber directed to the trap was estimated as follows. Firstly, the attenuation of the mechanical shutter was measured to be more than -50 dB, limited by photodetector noise. Secondly, no definite 397 nm light intensity dependence of the clock frequency outside measurement uncertainty was observed. We estimate the ac Stark shift due to 397 nm light from linear fitting of the clock frequency to the intensity of 397 nm light which was a zero measurement of 0.27(0.54) Hz at normal operation. A further experiment when the mechanical shutter was inoperative found that the blue shift of the clock transition due to leaked 397 nm radiation from the AOM was proportional to the incident intensity to the AOM. Within measurement uncertainty there was no difference in frequency between normal operation and when the shutter is inoperative and the 397 nm intensity is extrapolated to zero. This result excludes the possibility of 397nm light scattering from the back surface of the shutter blade and coupling to the fiber. No frequency shift is observed when the AOM for the clock laser is kept on and only a mechanical shutter is used to generate clock interrogation pulses, ruling out frequency chirp due to the transient phenomena of the AOM [24].

Ac Stark shift due to the 854-nm light (used for quenching the ion from the $^2D_{5/2}$ levels into the $^2S_{1/2}$ levels) was estimated from the comparison between frequencies measured with and without use of 854-nm light (when we do not use 854-nm light the ion in $^2D_{5/2}$ takes longer to decay via spontaneous emission.) We did not observe any frequency difference with uncertainty of 0.2 Hz. We also measured frequency shifts by changing the power of the clock laser and ac Stark shifts were evaluated to be 0 (0.3) Hz. The linear frequency drift of the clock laser is roughly compensated by a chirped AOM. The servo error was estimated from the analysis of the residual error, resulting in 0 (0.5) Hz. Summing the above, we calculate a total systematic shift of 4.2 Hz with an uncertainty of 0.9 Hz.

Table 1: Systematic shifts and their uncertainties of the $^{40}$Ca$^+$ clock

|  | Shift (Hz) | Uncertainty (Hz) | Evidence |
|---|---|---|---|
| Gravitational | 3.4 | 0.1 | Altitude measurement |
| Blackbody radiation | 0.39 | 0.05 | Temperature measurement of electrodes |
| Ac Stark due to 397nm | 0.27 | 0.54 | Measurement referring Sr clock (see text) |
| Ac Stark due to 854nm | 0 | 0.2 | Shift measurement referring Sr clock |
| Ac Stark due to clock laser | 0 | 0.3 | Shift measurement referring Sr clock |
| Electric quadrupole $\left(\left|^2S_{1/2}, M_J=1/2\right\rangle \rightarrow \left|^2D_{5/2}, M_J=3/2\right\rangle\right)$ | 0.1 | 0.3 | Measurement of six Zeeman components |
| Quadratic Doppler | 0 | <0.1 | Motional sidebands and no rf-photon correlation |
| Second-order Zeeman | 0 | <0.1 | Upper bound calculated from magnetic field bias of 3 μT |
| Stark due to secular motion | 0 | <0.01 | Motional sideband |
| Stark due to micromotion | 0 | <0.1 | Upper bound estimated from observed rf-photon correlation |
| Servo error | 0 | 0.5 | Residual error in locking to the atomic line and long-term drift rate of stray magnetic fields |
| Total | 4.2 | 0.9 |  |

### 3. Strontium lattice clock

The optical lattice clock utilized as a stable frequency reference in this work is based on the $^1S_0$–$^3P_0$ transition (λ=698 nm) of optically trapped spin-polarized fermionic $^{87}$Sr atoms ($^{87}$Sr lattice clock [9]). The absolute frequency of the $^{87}$Sr lattice clock at NICT agrees with other clocks in various institutes within the so-called Cs-limit [13]. The systematic uncertainty has been evaluated to be $5\times10^{-16}$. Using a 60 km fiber link [7] we confirmed the agreement of the clock frequencies at the $10^{-16}$ level with another $^{87}$Sr lattice clock in The University of Tokyo. This experiment also demonstrated that the instability of the clock at NICT reached $5\times10^{-16}$ in 1000 s [6]. Since then the instability of the clock laser has been improved by implementing optical compensation against vibration-induced phase noise, details of which are to be described elsewhere. By this technique the short term stability is $<2\times10^{-15}$ at 1 s and no longer requires the stability transfer technique employed previously [13]. Systematic shifts and their uncertainties have been recently evaluated after some minor renovations which has resulted in a systematic correction of -1.25(22) Hz. In this paper we report on the use of our lattice clock as an independent optical frequency reference for the evaluation of the Ca$^+$ clock.

### 4. Frequency ratio and instability

A Ti:Sapphire-based optical frequency comb [25] bridges the two clock transitions and the frequency ratio $\nu_{Ca}/\nu_{Sr}$ is measured as follows. The beat signal between the clock laser for the Sr lattice clock and the nearest comb component is first phase locked to a stable RF frequency $f_{PLL}$ which is generated by a direct digital synthesizer with reference to a hydrogen maser. In this case, the repetition frequency $f_{rep}$ is expressed as follows,

$$f_{rep} = \frac{\nu_{Sr} - f_{ceo} - f_{PLL}}{N_1},$$

where $N_1$ and $f_{ceo}$ are the mode number of the comb component and the carrier-envelop offset frequency of the comb. Measurement of the beat frequency $f_b$ between the Ca$^+$ clock frequency and the nearest comb component (mode number $N_2$) yeilds the transition frequency $\nu_{Ca}$ according to

$$\nu_{Ca} = f_{ceo} + N_2 f_{rep} + f_b.$$

The frequency ratio $\nu_{Ca}/\nu_{Sr}$ can be calculated as

$$\frac{\nu_{Ca}}{\nu_{Sr}} = \frac{N_2}{N_1} + \frac{(1 - N_2/N_1) f_{ceo} - (N_2/N_1) f_{PLL} + f_b}{\nu_{Sr}}$$

Here, the first term is the order of 1, whereas the second term is on the order of $10^{-7}$ since the numerator is a radio frequency of $\sim 10^7$ Hz and the denominator is an optical frequency of $\sim 10^{14}$ Hz. Therefore, the easily attainable $10^{-10}$ fractional accuracy of all components including $\nu_{Sr}$ is sufficient for evaluation of the frequency ratio at the $10^{-16}$ level.

The instability of $\nu_{Ca}/\nu_{Sr}$ is shown in Fig. 2. The level of instability slowly decreases with increasing integration time over 1-10 s, indicating our clock laser (employing a simple cylindrical shaped high finesse optical cavity) is slightly above the thermal noise limit. The optical compensation of vibration induced noise allows the Sr clock laser to operate at the thermal noise limit, but the compensation is not applied to the Ca$^+$ clock laser, which may prevent the short term stability from reaching the flat thermal noise limit. The long term instability of $2.4 \times 10^{-14}/\tau^{1/2}$, where $\tau$ is the integration time, is limited by the Ca$^+$ clock. The instability of the Sr lattice clock is $1 \times 10^{-14}/\tau^{1/2}$ according to an interleaved stabilizing operation, where the clock laser is stabilized to a π transition from an identical stretched state with a total operation cycle of 6 s. This is shown as the thin black trace in Fig. 2.

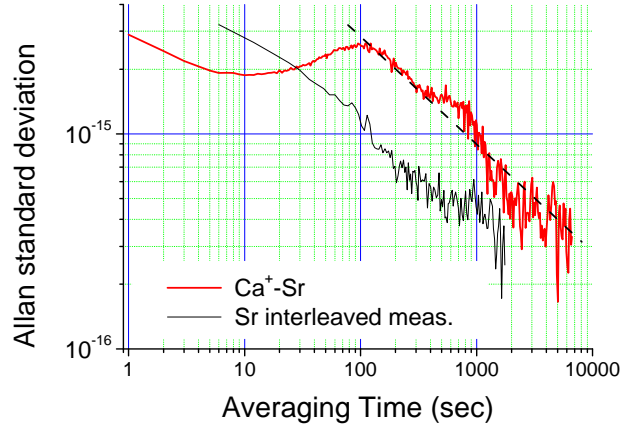

Fig.2 Instability of the frequency ratio $\nu_{Ca}/\nu_{Sr}$ (red) determined by the instability of Ca$^+$ clock. An interleaved measurement of the Sr lattice clock has resulted in lower instability as shown in black.

The $Ca^+$ center frequency is estimated by forty quantum projection measurements in a cycle time of 17 second with a Fourier-limited linewidth of 30 Hz. Random initial preparation of Zeeman substates ($m_J=\pm1/2$) halves the number of effective measurements. The quantum projection limit estimated with these parameters is $2\times10^{-14}/\tau^{1/2}$, and is consistent with the observed long term stability shown as a broken line in Fig.2. Note that the bump around averaging time of 800 s is suspected to be caused by two 10 m-length optical fiber cables that do not have fiber noise cancellation. According to the thermal coefficient of the refractive index of fused-silica, a peak-to-peak temperature fluctuations of 0.8 K are sufficient to induce this level of phase noise.

The frequency ratio $\nu_{Ca}/\nu_{Sr}$ measured for a half year are summarized in Fig.3. The results indicate that both optical clocks have a day-to-day reproducibility better than $1\times10^{-15}$, consistent with the evaluation of systematic shifts described above. Taking into account a statistical uncertainty of $1.8\times10^{-16}$ and systematic uncertainties of $Ca^+$ clock and Sr clock ($2.2\times10^{-15}$ and $5\times10^{-16}$ respectively), we conclude the frequency ratio $\nu_{Ca}/\nu_{Sr}$ is 0.957 631 202 358 049 9 with a fractional uncertainty of $2.3\times10^{-15}$.

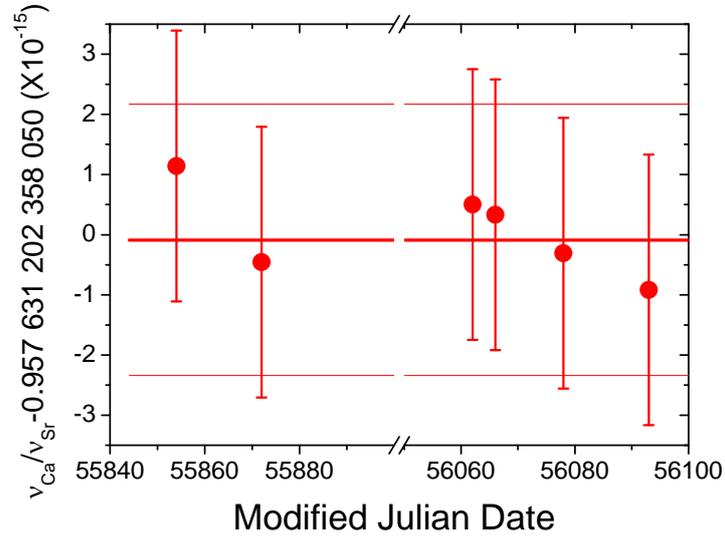

Fig.3 Frequency ratio $\nu_{Ca}/\nu_S$ obtained for half a year. The reproducibility of less than $10^{-15}$ is consistent with the systematic uncertainties of two clocks. The thick and thin lines indicate the weighted average and the uncertainty, respectively.

## 5. Absolute frequency measurement of the $Ca^+$ clock

The frequency of the $Ca^+$ clock transition was also evaluated by a microwave link to International Atomic Time (TAI). We measured a beat frequency between the clock laser and the nearest comb component of a frequency comb stabilized to a hydrogen maser. The frequency of the hydrogen maser is calibrated to UTC(NICT) using dual-mixer time difference (DMTD) technique [26]. Measurements spanning one year are summarized in Fig. 4. The link uncertainties consist of those in the UTC(NICT)-TAI link and TAI-TT link. In principle, these uncertainties are evaluated by BIPM in Circular T. However, the calibration of the UTC(NICT) to the TAI is based on time differences of two instances separated for 5 days, whereas the $Ca^+$ clock is operated only for several hours. We therefore conservatively estimate that UTC(NICT) for this limited time could be different from the five day average by the amount of the instability of UTC(NICT) over those 5 days. In the duration shown in Fig. 4, this 5-day instability of UTC(NICT) against TAI was $3.0\times10^{-15}$, where twelve data points based on nine calibrations of the UTC(NICT) is averaged to yield the final result. 45 days of signal integration (Nine calibrations) resulted in an the instability of TAI-UTC(NICT) of $1.9\times10^{-15}$. We suspect that the reduction is less than the expected factor of 3(ie. $9^{1/2}$) because of imperfect steering of UTC(NICT) to the ensemble of Cs clocks, TA(NICT). Thus, the uncertainty due to the TAI link is evaluated to be $1.9\times10^{-15}$ including minor contributions from the calibration of TAI to the SI second ($3\times10^{-16}$). Coupled with considerations of the systematic uncertainty of the $Ca^+$ clock shown in Table 1 and statistical uncertainty in Fig.4, the total fractional uncertainty of the absolute frequency measurement is $3.0\times10^{-15}$. We therefore determine that the absolute frequency for our measurement is 411 042 129 776 398.4(1.2) Hz.

The instability at the $10^{-16}$ level shown in Fig. 2 makes it possible to use the $Ca^+$ clock as a transfer oscillator, in the same way that a hydrogen maser is used in the microwave frequency regime. Combining the measured frequency ratio $\nu_{Ca}/\nu_{Sr}$ with the absolute frequency of our $Ca^+$ clock we evaluate the absolute frequency of the Sr lattice clock to be 429 228 004 229 871.9 (1.2) Hz. This frequency is consistent with our previous measurement [13] as well as that obtained in four other institutes [10-12, 27], indicating the validity of the TAI-link we have used.

While our $Ca^+$ clock has a day-to-day reproducibility at the $10^{-16}$ level and the absolute frequency agrees with the CIPM recommendation [28], there is a discrepancy between two other measurements [14, 15]. To investigate this discrepancy we are planning some further experiments; firstly we will in the near future make direct international frequency comparison via GPS carrier-phase techniques, and secondly we will repeat the frequency measurement using an apparatus of an $In^+-Ca^+$ clock currently under development [29].

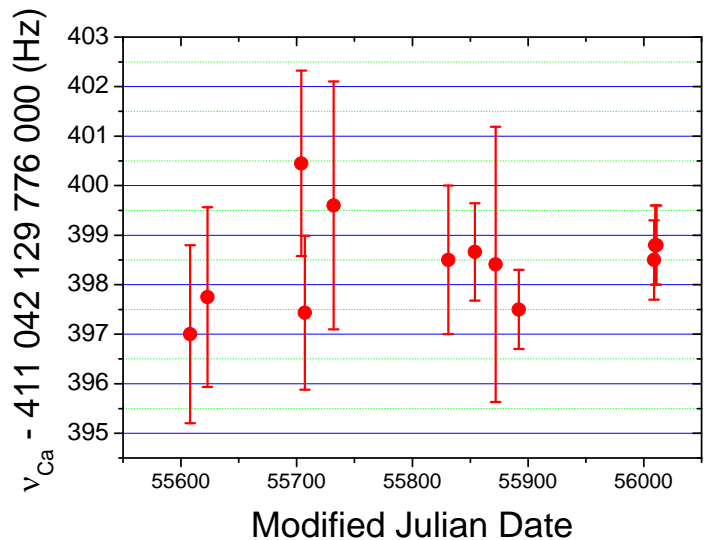

**Fig.4** One year record of absolute frequency measurements of $^{40}Ca^+$ clock frequency. The corrections due to systematic shifts are included. Error bars, however, include only statistical errors. The hydrogen-maser frequency used for the frequency-comb stabilization was calibrated using the TAI link. Calibrations of UTC(NICT) of two data at MJD = 55704 and 55707 as well as three data at MJD=56009-56011 are based on same Circular T. The weighted average frequency and statistical uncertainty is 411 042 129 776 398.4 (0.2) Hz.

## 6. Summary and Outlook

We have measured the frequency ratio between the $^{40}$Ca$^+$ $^2S_{1/2}$-$^2D_{5/2}$ transition and the $^{87}$Sr $^1S_0$-$^3P_0$ transitions with an uncertainty of $2.3 \times 10^{-15}$. The frequency ratio measurement demonstrated for the first time that the Ca$^+$ clock is able to reach the $10^{-16}$ level of instability. All optical comparison has also enabled quick and rigorous evaluation of the systematic shifts of Ca$^+$ ion clock. The frequency ratio $\nu_{Ca} / \nu_{Sr}$ was consistent with the absolute frequency measurement of the Ca$^+$ clock.

The Ca$^+$ and Sr system investigated here are two of the most popular optical clocks in ion-based and lattice-based systems. The systematic uncertainty of the Ca$^+$ clock in this work is at the $10^{-15}$ level, and we have demonstrated it may be used as a transfer oscillator. The systematic uncertainties of both clocks are currently projected to be smaller than current state-of-the-art cesium fountain clocks and therefore further improved measurement of the ratio $\nu_{Ca} / \nu_{Sr}$ may become an infrastructure of frequency metrology, enabling frequency comparison of two physically separated optical clocks more precisely than that of comparisons via SI second. In this case one optical clock can be locally linked to a Sr lattice clock and the remote clock linked to a Ca$^+$ clock possibly working as a logic ion in quantum logic gate spectroscopy. Such comparison requires only 1000 seconds of frequency ratio measurement in each laboratory.

## Acknowledgements

We are grateful to R. Kojima and A. Yamaguchi for considerable effort to initiate the clock operation of the Ca$^+$ clock and the Sr lattice clock, respectively. Contributions of Y. Hanado, H. Ito, and M. Kumagai are gratefully acknowledged in discussions related to the TAI-link. The authors thank Y. Koyama and N. Shiga for fruitful discussions. Experimental support by H. Ishijima and K. Kido were indispensable for this work. This research was supported in part by the Japan Society for the Promotion of Science through its FIRST program.